\documentclass[runningheads]{llncs}
\usepackage[T1]{fontenc}
\usepackage{balance}
\usepackage{graphicx}
\usepackage{multirow}
\usepackage{multicol}
\usepackage{float}
\usepackage{xcolor}
\usepackage{amssymb}
\usepackage{amsmath}
\usepackage{minibox}
\usepackage{tabularx}
\usepackage{caption}
\usepackage{subcaption}
\usepackage{pgf}
\usepackage{enumitem}
\usepackage{array}

\usepackage{listings}
\newcommand{\cylon}{\textit{Cylon}}

\newcommand{\gcylon}{\textit{GCylon}}

\begin{document}

\title{High Performance Dataframes from Parallel Processing Patterns}

\author{
Niranda Perera\inst{1}\orcidID{0000-0003-3076-0011} \and
Supun Kamburugamuve\inst{2} \and
Chathura Widanage\inst{2} \and
Vibhatha Abeykoon\inst{2} \and
Ahmet Uyar\inst{2} \and
Kaiying Shan\inst{3} \and
Hasara Maithree\inst{4} \and
Damitha Lenadora\inst{5} \and
Thejaka Amila Kanewala\inst{2} \and
Geoffrey Fox \inst{6} 
}

\authorrunning{Perera et al.}

\institute{
Luddy School of Informatics, Computing, and Engineering, Indiana University, Bloomington, IN 47408, USA \and
Indiana University Alumni, Bloomington, IN 47405, USA \and
University of Virginia, Charlottesville, VA 22904, USA \and
University of Moratuwa, Bandaranayake Mawatha, Moratuwa 10400, Sri Lanka \and
University of Illinois Urbana-Champaign,  Urbana, IL 61801, USA \and
Biocomplexity Institute and Initiative, University of Virginia, Charlottesville, VA 22904, USA
}

\maketitle

\begin{abstract}

The data science community today has embraced the concept of \textit{Dataframes} as the de facto standard for data representation and manipulation. Ease of use, massive operator coverage, and popularization of R and Python languages have heavily influenced this transformation. However, most widely used serial Dataframes today (R, \texttt{pandas}) experience performance limitations even while working on even moderately large data sets. We believe that there is plenty of room for improvement by investigating the generic distributed patterns of dataframe operators.
 


In this paper, we propose a framework that lays the foundation for building high performance distributed-memory parallel dataframe systems based on these parallel processing patterns. We also present \cylon, as a reference runtime implementation. We demonstrate how this framework has enabled \cylon{} achieving scalable high performance. We also underline the flexibility of the proposed API and the extensibility of the framework on different hardware.  To the best of our knowledge, \cylon{} is the first and only distributed-memory parallel dataframe system available today.

\end{abstract}


\keywords{Dataframes \and High performance computing \and Data engineering \and Relational algebra \and MPI \and Distributed Memory Parallel}


%

\section{Introduction}

The Data Science domain has expanded monumentally in both research and industry communities over the past few decades, predominantly owing to the \textit{Big Data} revolution. Artificial Intelligence (AI) and Machine Learning (ML) offer even more complexities to data engineering applications, which are now required to process terabytes of data. Typically, a significant amount of \textit{developer time} is spent on data exploration, preprocessing, and prototyping while developing AI/ML pipelines. Therefore, improving its efficiency directly impacts the overall pipeline performance.



With the wide adoption of R and Python languages, the data science community is increasingly moving away from established SQL-based abstractions. \textit{Dataframes} play a pivotal role in this transformation \cite{mckinney2011pandas} by providing a functional interface and interactive development environment for exploratory data analytics. \texttt{pandas} is undoubtedly the most popular dataframe library available today. 
Its open source community has grown significantly, and the API has expanded up to 200+ operators. Despite this popularity, both R-dataframe and pandas encounter performance limitations even on moderately large data sets. In our view, dataframes have now exhausted the capabilities of a single computer, which paves way for distributed dataframe systems. 

There are several significant engineering challenges related to developing a scalable and high performance distributed dataframe system (Section \ref{sec:challenges}).
In this paper, we analyze dataframe operators to establish a set of generic distributed operator patterns and present an open-source high performance distributed dataframe system framework based on them, \cylon{}. We take inspiration from Mattson et al's \textit{Patterns for Parallel Programming} \cite{Mattson2004Patterns}. Our main focus is to present a mechanism that promotes an existing serial/ local operator into a distributed operator (Section \ref{sec:sys-cons}, \ref{sec:model}). The proposed framework is aimed at a distributed memory system executing in a Bulk Synchronous Parallel (BSP) \cite{valiant1990bridging,fox1989solving} environment. This combination has been widely employed by the high performance computing (HPC) community for exascale computing applications with admirable success. 

\section{Dataframe Systems}
\label{sec:df}


A \textit{dataframe} is a heterogeneous data structure containing a set of arrays that are individually homogeneous. 
In contrast, deep learning or machine learning use \textit{tensors} which are homogeneously typed multidimensional arrays. These two data structures are integrated to support end-to-end data engineering workloads. 
Dataframes were first introduced by the S language in 1990, and their popularity grew exponentially with R and Python languages\cite{mckinney2011pandas}. 
These libraries contain a large number of SQL-like statistical, linear algebra and, relational algebra operators and are sequential in execution. With the increasing size of data, there have been some attempts to scale dataframe execution both in the cloud and high performance computing environments such as, Dask\cite{rocklin2015dask}, Modin\cite{petersohn2020towards}, and Koalas. 


\subsection{Engineering Challenges}
\label{sec:challenges}


While there is a compelling need for a distributed dataframe system, there are several engineering challenges.

\begin{itemize}[wide=0pt]

    \item \textbf{Lack of Specification}:
    Despite the popularity, there is very little consensus on a specification/standard for dataframes and their operators amongst the systems available today. 
Rapid expansion in applications and the increasing demand for features may have contributed to this divergence. 
The current trend is to use \texttt{pandas} as the reference API specification \cite{petersohn2020towards}, and we also follow this approach for the work described in this paper. 

\item \textbf{Massive API}: \texttt{pandas} API consists of ~240 operators \cite{abeykoon2021hptmt,petersohn2020towards}. There is also significant redundancy amongst the operators. 
It would be a mammoth undertaking to parallelize each of these operators individually. Petersohn et al \cite{petersohn2020towards}, have taken a more practical approach by identifying a core set of operators (\textit{Dataframe Algebra}) listed in Table \ref{tab:modin-ops}. 
In this paper, we have taken a different approach by identifying distributed patterns in dataframe operators, and devise a framework that can best scale them in a distributed memory parallel environment.


\vspace{1em}
\begin{minipage}{\linewidth}
\begin{minipage}[c]{0.64\linewidth}
\centering
\includegraphics[width=\linewidth]{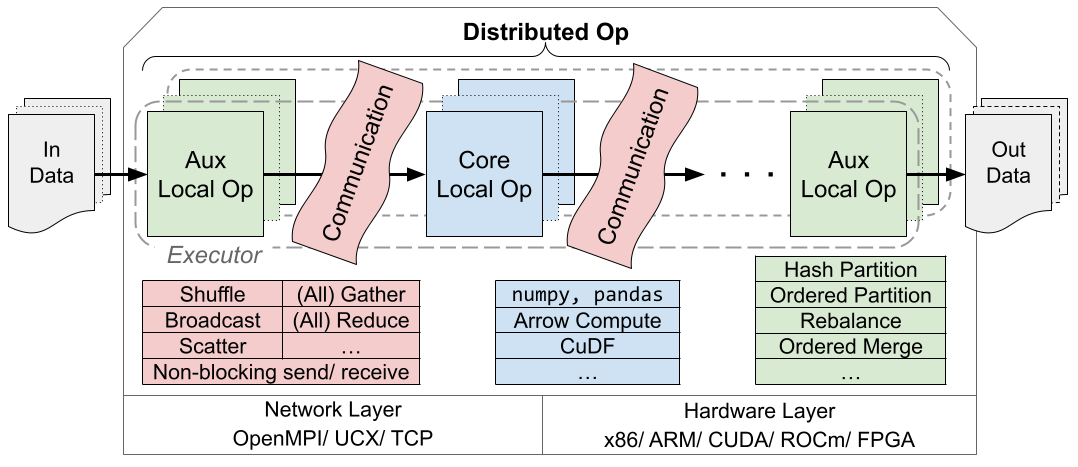}
\captionsetup{justification=centering,skip=0pt}
\captionof{figure}{Distributed Memory Dataframe Abstraction}
\label{fig:dist-op}
\end{minipage}
\hfill
\begin{minipage}[c]{0.34\linewidth}
\centering
\scriptsize
\begin{tabular}{|>{\centering\arraybackslash}p{0.4\linewidth}>{\centering\arraybackslash}p{0.4\linewidth}|} 
\hline
Selection  & Window       \\ 
Projection & Transpose    \\ 
Union      & Map          \\ 
Difference & Aggregation$^{*}$\\ 
Join       & ToLabels     \\ 
Unique     & FromLabels   \\ 
GroupBy    & Rename       \\ 
Sort       & \\
\hline
\multicolumn{2}{c}{\small *Not categorized in Modin} 
\end{tabular}
\captionsetup{justification=centering,skip=0pt}
\captionof{table}{Modin DataFrame Algebra \cite{petersohn2020towards}}
\label{tab:modin-ops}
\end{minipage}
\end{minipage}
\vspace{1em}

\item \textbf{Efficient Parallel Execution}: Distributed data engineering systems generally vary in their execution model. Dask, Modin, and Koalas dataframes are built on top of a fully asynchronous execution environment. 
Conversely, Bulk-Synchronous-Parallel (BSP) model is used in data parallel deep learning. This mismatch poses a challenge in creating a fully integrated scalable data engineering pipeline. Our framework attempts to bridge this gap by taking an HPC approach to parallelizing Dataframe operators. 

\end{itemize}






\subsection{System Considerations}
\label{sec:sys-cons}

There are multiple aspects that need to be considered when developing a distributed data processing framework \cite{kamburugamuve2021hptmt}. Our distributed dataframe model is designed based on the following considerations.

\begin{itemize}[wide=0pt]



\item \textbf{BSP Execution}: 
The most widely used \textbf{execution models} are, 1) \textit{Bulk Synchronous Parallel} \cite{valiant1990bridging,fox1989solving} and 2) \textit{Fully Asynchronous}. The former assumes all the tasks are executing in parallel, and the executors synchronize with each other by exchanging messages at certain points. The sections of code between communication synchronizations execute independently. In the latter, tasks would be executed independently. Input and output messages will be delivered using queues, and often this requires a central scheduler to orchestrate the tasks. 
Many recent data engineering frameworks (e.g. Apache Spark, Dask, etc.) have adopted fully asynchronous execution. Our framework is based on BSP execution in a distributed memory environment. Gao et al \cite{gao2021scaling} recently published a similar concept for scaling joins over thousands of GPUs. We intend to show that this approach generalizes to all operators and achieves commendable scalability and high performance.

\item \textbf{Distributed Memory}: Most often the parallel \textbf{memory model} of a system is a choice between,
1) \textit{Shared}: multiple CPU cores in a single machine via threads/ processes (e.g. OpenMP), 2) \textit{Distributed}: every instance of the program is executed on an isolated memory, and data is communicated via message passing (e.g. MPI), and 3) \textit{Hybrid}: combines shared and distributed models. Our framework is developed based on Distributed memory. 








\item \textbf{Columnar Data Format}:  
Most of dataframe operators access data along columns, and using a columnar format allows operators to be vectorized using SIMD and hardware accelerators (e.g. GPUs). As a result, the patterns described in this paper focus on columnar dataframes. 




\item \textbf{Row-based Partitioning}: 
Dataframe partitioning is semantically different from traditional matrix/tensor partitioning. Due to the homogeneously typed data storage, when a matrix/ tensor is partitioned, the effective computation reduces for each individual partition. 
By comparison, dataframe operator patterns (Section \ref{sec:op-pat}) show that not all columns of a dataframe contribute equally to the computation, e.g. \texttt{join} is performed on \textit{key} columns, while the rest of the columns move alongside the keys. Both Apache Spark \cite{apache-spark} and Dask \cite{rocklin2015dask} follow a row-based partitioning scheme, while Modin \cite{petersohn2020towards} uses block-based partitioning with dynamic partition ID allocation. 
Our framework employs BSP execution on a distributed memory parallel environment. We would like to distribute the computation among all available executors to maximize the scalability. We also use row-based partitioning because it allows us to hand over the data partitions with identical schema to each executor.

\end{itemize}








\section{Distributed Memory Dataframe Framework} 
\label{sec:model}

The lack of a specification presents a challenge in properly defining a \textit{dataframe} data structure. It is not quite a relation in an SQL sense, nor a matrix/multidimensional array. For our distributed memory model, we borrow definitions from Petersohn et al \cite{petersohn2020towards}. 
Dataframes contain heterogeneously typed data originating from a known set of \textit{domain}s, $Dom = \{dom_1, dom_2, ...\}$. For dataframes, these \textit{domain}s represent all the data types they support. 

\begin{definition}\label{def:schema}
  A \textbf{Schema} of a Dataframe, $S_M$ is a tuple $(D_M, C_M)$, where $D_M$ is a vector of $M$ domains and $C_M$ is a vector of $M$ corresponding column labels. Column labels usually belong to \textit{String}/ \textit{Object} domain. 
\end{definition}

\begin{definition}\label{def:df}
  A \textbf{Dataframe} is a tuple $(S_M, A_{NM}, R_N)$, where $S_M$ is the Schema with $M$ domains, $A_{NM}$ is a 2-D array of entries where actual data is stored, and $R_N$ is a vector of $N$ row labels belonging to some domain. \emph{Length} of the Dataframe is $N$, i.e. the number of rows.
\end{definition}

\subsection{Distributed Memory Dataframe}
\label{sec:dist-df}


\textit{"How to develop a high performance scalable dataframe runtime?"} is the main problem we aim to address in our framework.
We attempt to promote an already available \textit{serial (local) operator} into a distributed-memory parallel execution environment (Figure. \ref{fig:dist-op}). For this purpose, we extend the definition of a dataframe for a distributed memory parallel execution environment with row-based partitioning. 

\begin{definition}\label{def:ddf}
  A \textbf{Distributed-Memory Dataframe} (DMDF) is a \emph{virtual} collection of $P$ \emph{Dataframes} (named \emph{Partition}s) of lengths $\{N_0,...,N_{P-1}\}$ and a common Schema $S_M$. Total length of the DMDF is $\Sigma N_i = N$, and the row labels vector is the concatenation of individual row labels, $R_N=\{R_0R_1...R_{P-1}\}$.
\end{definition}

\begin{minipage}{0.99\linewidth}
\begin{minipage}[c]{0.49\linewidth}
\centering
\includegraphics[width=0.9\linewidth]{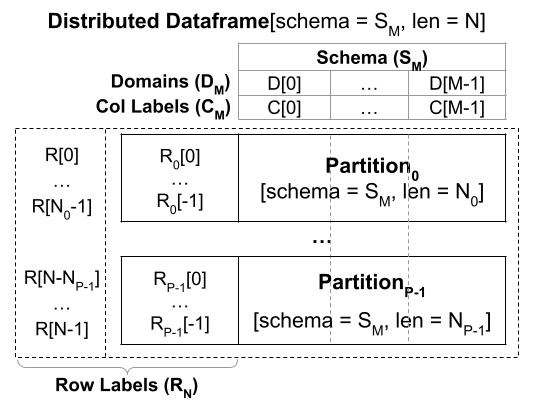}
\captionsetup{justification=centering,skip=0pt}
\captionof{figure}{Distributed Memory Dataframe}
\label{fig:dist-df}
\end{minipage}
\hfill
\begin{minipage}[c]{0.49\linewidth}
\centering
\scriptsize
\begin{tabular}{c|ccc|}
\cline{2-4}
\multicolumn{1}{l|}{}                    & \multicolumn{3}{c|}{\textbf{Data Structure}}                                                \\ \hline
\multicolumn{1}{|c|}{\textbf{Operation}} & \multicolumn{1}{c|}{\textbf{Dataframe}} & \multicolumn{1}{c|}{\textbf{Array}} & \textbf{Scalar} \\ \hline
\multicolumn{1}{|c|}{Shuffle (AllToAll)} & \multicolumn{1}{c|}{Common}         & \multicolumn{1}{c|}{Rare}           & N/A             \\ \hline
\multicolumn{1}{|c|}{Scatter}            & \multicolumn{1}{c|}{Common}         & \multicolumn{1}{c|}{Rare}           & N/A             \\ \hline
\multicolumn{1}{|c|}{Gather/AllGather}  & \multicolumn{1}{c|}{Common}         & \multicolumn{1}{c|}{Common}         & Common          \\ \hline
\multicolumn{1}{|c|}{Broadcast}          & \multicolumn{1}{c|}{Common}         & \multicolumn{1}{c|}{Common}         & Common          \\ \hline
\multicolumn{1}{|c|}{Reduce/AllReduce}  & \multicolumn{1}{c|}{N/A}            & \multicolumn{1}{c|}{Common}         & Common          \\ \hline
\end{tabular}
\captionsetup{justification=centering,skip=0pt}
\captionof{table}{Communication semantics in Dataframe Operators and the frequency of occurrence}
\label{tab:com-semantics}
\end{minipage}
\end{minipage}

\subsection{\textbf{Building Blocks}}
\label{sec:components}
As shown in Figure \ref{fig:dist-op}, a distributed operator is comprised of multiple components/ building blocks, such as,

\begin{enumerate}[wide=0pt]
\item \textbf{Data Structures}: The distributed memory framework we employ uses three main data structures: dataframes, arrays, and scalars. While most of the operators are defined on dataframes, arrays and scalars are also important because they present different communication semantics.





\item \textbf{Serial/Local Operators}: These refer to single-threaded implementations of core operators (Table \ref{tab:modin-ops}). There could be one or more libraries that provide this functionality (e.g. \texttt{numpy}, \texttt{pandas}, RAPIDS CuDF, Apache Arrow Compute, etc). Choice of the library depends on the language runtime, the underlying memory format, and the hardware architecture. 


\item \textbf{Communication Routines}: A BSP execution allows the program to continue independently until the next communication boundary is reached (Section \ref{sec:sys-cons}). HPC message passing libraries such as MPI (OpenMPI, MPICH, MSMPI) and UCX provide communication routines for memory buffers (works for homogeneously typed arrays). The most primitive routines are tag-based \textit{async send} and \textit{async receive}. Complex patterns (generally termed \textit{collectives}) can be derived on top of these two primitive routines (e.g. MPI-Collectives, UCX-UCC). 
The columnar data format represents a column by a tuple of buffers 
and a dataframe is a collection of such columns. Therefore, a communication routine would have to be called on each of these buffers. 
We identified a set of communication routines required to implement distributed memory dataframe operators. These are listed in Table \ref{tab:com-semantics}. 

\item \textbf{Auxiliary Operators}: \textit{Partition} operators are essential for distributed memory applications. Partitioning determines how a local data partition is split into subsets so that they can be sent across the network. This operator is closely tied with \textit{Shuffle} communication routine. The goal of \textit{hash partitioning} is to assign a partition ID to each row of the dataframe so that at the end of the communication routine, all the equal/key-equal rows end up in the same partition. \emph{Ordered Partitioning} is used when the operators (e.g. \textit{Sort}) need to be arranged based on sorted order. 
Parallel sorting on multiple key-columns further complicates the operation by accessing values along row-dimension (cache-unfriendly). 
\emph{Rebalance} repartitions data across the executors equally or based on a sequence of rows per partition. 
On average, an executor would only have to exchange data with its closest neighbors to achieve this. To determine the boundaries, the executors must perform an \textit{AllGather} on their partition lengths. 
\textit{Merge} is another important auxiliary operator. It is used to build the final ordered dataframe in \textit{Sort} operator to merge individually ordered sub-partitions ($\sim$merge-sort).

\end{enumerate}

\subsection{\textbf{Generic Operator Patterns}}
\label{sec:op-pat}
\begin{table}[htbp]
\vspace{-2em}
\scriptsize
\centering
\begin{tabular}{|l|c|c|c|} 
\hline
\multicolumn{1}{|c|}{\textbf{Pattern}}                                                                         & \textbf{Operators}                                                              & \begin{tabular}[c]{@{}c@{}}\textbf{Result }\\\textbf{Semantic}\end{tabular} & \textbf{Communication}                                                       \\ 
\hline
\textbf{Embarrassingly parallel}                                                                                        & \begin{tabular}[c]{@{}c@{}}Select, Project, Map, \\Row-Aggregation\end{tabular} & Partitioned                                                                 & -                                                                            \\ 
\hline
\textbf{Loosely Synchronous}                                                                                            & \multicolumn{1}{l|}{}                                                           & \multicolumn{1}{l|}{}                                                       & \multicolumn{1}{l|}{}                                                        \\
\begin{tabular}{@{\labelitemi\hspace{\dimexpr\labelsep+0.5\tabcolsep}}l@{}}Shuffle Compute\end{tabular}        & \begin{tabular}[c]{@{}c@{}}Union, Difference, \\Join, Transpose\end{tabular}    & Partitioned                                                                 & Shuffle                                                                      \\
\begin{tabular}{@{\labelitemi\hspace{\dimexpr\labelsep+0.5\tabcolsep}}l@{}}Combine Shuffle Reduce\end{tabular} & Unique, GroupBy                                                                 & Partitioned                                                                 & Shuffle                                                                      \\
\begin{tabular}{@{\labelitemi\hspace{\dimexpr\labelsep+0.5\tabcolsep}}l@{}}Broadcast Compute\end{tabular}      & Broadcast-Join$^{*}$                                                            & Partitioned                                                                 & Bcast                                                                        \\
\begin{tabular}{@{\labelitemi\hspace{\dimexpr\labelsep+0.5\tabcolsep}}l@{}}Globally Reduce\end{tabular}        & Column-Aggregation                                                              & Replicated                                                                  & AllReduce                                                                    \\
\begin{tabular}{@{\labelitemi\hspace{\dimexpr\labelsep+0.5\tabcolsep}}l@{}}Globally Ordered\end{tabular}       & Sort                                                                            & Partitioned                                                                 & \begin{tabular}[c]{@{}c@{}}Gather, Bcast, Shuffle, AllReduce\end{tabular}  \\
\begin{tabular}{@{\labelitemi\hspace{\dimexpr\labelsep+0.5\tabcolsep}}l@{}}Halo Exchange\end{tabular}          & Window                                                                          & Partitioned                                                                 & Send-recv                                                                    \\ 
\hline
\textbf{Partitioned I/O}                                                                                          & Read/Write                                                                      & Partitioned                                                                 & \begin{tabular}[c]{@{}c@{}}Send-recv, Scatter, Gather\end{tabular}         \\ 
\hline
\multicolumn{4}{c}{*Specialized join algorithm}          \end{tabular}
\caption{Generic Dataframe Operator Patterns}
\label{tab:op-patterns}  
\vspace{-3em}
\end{table}


Our key observation is that dataframe operators can be categorized into several generic parallel execution patterns. We believe a distributed framework based on these patterns would make the parallelization of the massive API more tractable. These generic patterns (Table \ref{tab:op-patterns}) have distinct distributed execution semantics, and individually analyzing the semantics allowed us to recognize opportunities for improvement. 
Rather than optimizing each operator individually, we can focus more on improving bottlenecks of the pattern, and thereby benefiting all operators derived from it.

\textbf{Result Semantic}: A local dataframe operator may produce dataframes, arrays, or scalars as results. When we promote a local operator to distributed memory, these result semantics could be nuanced (a global-viewed dataframe). Distributed memory dataframes (and arrays) are partitioned, and therefore a dataframe/array result (e.g. \texttt{select, join, etc.}) should also be partitioned. By contrast, scalars cannot be partitioned, so when an operator produces a scalar, it needs to be \textit{replicated} to preserve the overall operator semantic. 



\subsubsection{Embarrassingly Parallel (EP)}
\label{sec:em-par}

EP operators are the most trivial class of operators. They do not require any communication to parallelize the computation. \textit{Select}, \textit{Project}, \textit{Map}, and \textit{Row-Aggregation} fall under this pattern. While \textit{Select} and \textit{Map} apply to rows, \textit{Project} works by selecting a subset of columns. These operations are expected to show linear scaling. 
Arithmetic operations (e.g. \texttt{add}, \texttt{mul}, etc.) are good examples of this pattern. 

\subsubsection{Loosely Synchronous}

\begin{enumerate}[wide=0pt]
    \item \textbf{Shuffle-Compute}: This is a common pattern that can be used for operators that depend on \textit{Equality/Key Equality of rows}. Of the core dataframe operators, \texttt{join, union} and \texttt{difference} directly fall under this pattern, while \texttt{transpose} follows a more nuanced approach. 

Hash partitioning and shuffle communication rearrange data in such a way that equal/key-equal rows are on the same partition. Corresponding local operation can then be called trivially. \textit{Join, Union} and \textit{Difference} operators follow this pattern: 

{\centering\scriptsize
\fcolorbox{black}{white}{HashPartition}$\rightarrow$\fcolorbox{black}{lightgray}{Shuffle}$\rightarrow$\fcolorbox{black}{white}{LocalOp}
\par}
The local operator may access memory randomly, and allowing it to work on in-cache data improves the efficiency of the computation. We could also simply attach a \textit{local hash partition} block at the end of the shuffle to achieve this since hash-partitioning can stream along the columnar data and is fairly inexpensive. 

{\centering\scriptsize
\fcolorbox{black}{white}{HashPartition}$\rightarrow$\fcolorbox{black}{lightgray}{Shuffle}$\rightarrow$
\fcolorbox{black}{white}{LocalHashPartition}$\rightarrow$\fcolorbox{black}{white}{LocalOp}
\par}
A more complex scheme would be to hash-partition data into much smaller sub-partitions from the start. Possible gains on each of these schemes depend heavily on runtime characteristics.

\textit{Transpose} is important for dataframe \textit{Pivot} operations. It can be implemented without communication in a block partitioned environment \cite{petersohn2020towards}. In a row partitioned setup, a \textit{shuffle} is required at the end of block-wise local transpose to rearrange the blocks. 



\item \textbf{Combine-Shuffle-Reduce}: An extension of the \textit{Shuffle-Compute} pattern, Combine-Shuffle-Reduce is semantically similar to the famous MapReduce paradigm. The operations that reduce the resultant dataframe length such as \textit{Groupby} and \textit{Unique}, could benefit from this pattern.
The initial local operation would reduce data into a set of intermediate results (similar to the combine step in \textit{MapReduce}) e.g. \texttt{groupby.std}, creating \texttt{sum$\_x^2$}, \texttt{sum\_$x$}, and \texttt{count\_$x$}, which would then be shuffled. Upon their receipt, a local operation is performed to finalize the results. Perera et al \cite{perera2020fast} also discuss a similar approach for dataframe reductions. The effectiveness of \textit{combine-shuffle-reduce} over \textit{shuffle-compute} depends on the \textit{Cardinality} $(\mathbf{C})$ (Section \ref{sec:run_asp}). 

{\centering\scriptsize
\minibox[frame, c]{LocalOp (interm. res.)}$\rightarrow$\minibox[frame, c]{HashPartition}$\rightarrow$\fcolorbox{black}{lightgray}{Shuffle}$\rightarrow$\minibox[frame, c]{LocalOp (final res.)}
\par}

\item \textbf{Broadcast-Compute}: This requires a \textit{broadcast} routine rather than \textit{shuffle}. \texttt{broadcast\_join}, a special algorithm for join, is a good example of this pattern. Broadcasting the smaller length relation to all other partitions and performing a local join is potentially much more efficient than shuffling both relations. 

\item \textbf{Globally-Reduce}: This is most commonly seen in dataframe \textit{Column Aggregation} operators. It is similar to EP, but requires communication to arrive at the final result. For example, calculating the column-wise \texttt{mean} requires a local summation, a global reduction, and a final value calculation. Some utility methods such as \textit{distributed length} and \textit{equality} also follow this pattern. For large data sets, the complexity of this operator is usually governed by the computation rather than the communication.

{\centering\scriptsize
\fcolorbox{black}{white}{LocalOp}$\rightarrow$
\fcolorbox{black}{lightgray}{Allreduce}$\rightarrow$
\fcolorbox{black}{white}{Finalize}
\par}

\item \textbf{Halo Exchange}: This is closely related to window operations. \texttt{pandas} API supports rolling and expanding windows. For row-partitions, the windows at the boundaries would have to communicate with their neighboring partitions and exchange partially computed results. The amount of data sent/received is based on the window type and individual length of partitions. 

\item \textbf{Globally Ordered}: Ascending order of rows $(row_i\leq row_j)$ holds if all elements in $row_i$ are less than or equal to the corresponding element in $row_j$. 
\textit{Ordered partitioning} preserves this order along the partition indices. For a single numerical key-column, the data can be range-partitioned based on a key-data histogram. 

{\centering\scriptsize
\minibox[frame, c]{Sample}$\rightarrow$
\fcolorbox{black}{lightgray}{\begin{tabular}[m]{@{}c@{}}Allreduce range\end{tabular}}$\rightarrow$
\minibox[frame, c]{Range part.}$\rightarrow$
\fcolorbox{black}{lightgray}{Shuffle}$\rightarrow$
\minibox[frame, c]{Local sort}
\par}

For multiple key-columns, we use \textit{sample sort} with regular sampling \cite{li1993versatility}. It sorts data locally and sends out a sample to a central entity that determines pivot points for data. Based on these points, sorted data will be split and shuffled, and finally all executors merge the received sub-partitions locally. 

{\centering\scriptsize
\minibox[frame, c]{Local\\sort}$\rightarrow$
\minibox[frame, c]{Sample}$\rightarrow$
\fcolorbox{black}{lightgray}{\begin{tabular}[m]{@{}c@{}}Gather\\@rank0\end{tabular}}$\rightarrow$
\minibox[frame, c]{Calc. pivots\\@rank0}$\rightarrow$
\fcolorbox{black}{lightgray}{\begin{tabular}[m]{@{}c@{}}Bcast\\pivots\end{tabular}}$\rightarrow$
\minibox[frame, c]{Split}$\rightarrow$
\fcolorbox{black}{lightgray}{Shuffle}$\rightarrow$
\minibox[frame, c]{Local\\merge}
\par}



\subsubsection{Partitioned I/O} 
\label{sec:io}
\textit{Partitioned Input} parallelizes the input data (CSV, JSON, Parquet) by distributing the files to each executor. It may distribute a list of input files to each worker evenly. Alternatively, it receives a custom one-to-many mapping from worker to input file(s) and reads the input files according to the custom assignment. 
In \textit{Partitioned Output}, each executor writes its own partition dataframe to one file.

\end{enumerate}



\subsection{Runtime Aspects}
\label{sec:run_asp}

\begin{itemize}[wide=0pt]



\item \textbf{Cardinality}: Hash-shuffle in \textit{Shuffle-Compute} pattern roughly takes $O(n)+O(\log_{}P*n)$, where $n$ is average length of a partition. In the \textit{Combine-Shuffle-Reduce} pattern, the initial local operation has the potential to reduce communication order to $n^{\prime} < n$. This gain depends on the \textit{Cardinality} ($\mathbf{C}$) of the dataframe $\mathbf{C} \in [\frac{1}{N}, 1]$, which is the number of unique rows relative to the length. $\mathbf{C} \sim \frac{1}{N} \implies n^{\prime} \lll n$, making the combine-shuffle-reduce much more efficient than a shuffle-compute. Consequently, when $\mathbf{C} \sim 1 \implies n^{\prime} \sim n$ may in fact worsen the combine-shuffle-reduce complexity. In such cases, shuffle-compute pattern is more efficient (\ref{sec:exper}). 

\item \textbf{Data Distribution}: This heavily impacts the partitioning operators. 
When there are unbalanced partitions, some executors may be underutilized, thereby affecting the overall distributed performance. \textit{Work-stealing} scheduling is a possible solution to this problem. In a BSP environment, pseudo-work-stealing execution can be achieved by storing partition data in a shared object store. 
Some operations could employ different operator patterns based on the data distribution. (e.g. When one relation is very small, \textit{Join} could use a \texttt{broadcast\_join}).



\item \textbf{Logical Plan Optimizations}:
An application consists of multiple Dataframe operator. Semantically, they are arranged in a DAG (directed acyclic graph), i.e. \textit{logical plan}. An \textit{optimized logical plan} can be generated based on rules (e.g. predicate push-down) or cost metrics. While these optimizations produce significant gains in real-life applications, this is an orthogonal detail to the individual operator patterns we focus on in this paper.

\end{itemize}

\section{\cylon}
\label{sec:cylon}


\cylon\ is a reference distributed memory parallel dataframe runtime based on Section \ref{sec:model}. 
We extended concept to implement a similar GPU Dataframe system, \gcylon. The source code is openly available in GitHub \cite{cylon-git} under Apache License.

\subsection{Architecture}

\begin{itemize}[wide=0pt]
    \item \textbf{Arrow Format \& Local Operators}: 
\cylon\ was developed in C++ using Apache Arrow Columnar format, which allows zero-copy data transfer between language runtimes. 
Arrow C++ Compute library is used for the local operators where applicable. Some operators were developed in-house.
Additionally, we use \texttt{pandas} and \texttt{numpy} in Python for EP operators. 

\item \textbf{Communication}:
\cylon\ currently supports MPI (OpenMPI, MPICH, MSMPI), UCX, and Gloo communication frameworks. The communication routines (Table \ref{tab:com-semantics}) are implemented using a collection of non-blocking routines on internal dataframe buffers. For the user, it would be a blocking routine on dataframes. For example, \textit{Dataframe Gather} is implemented via a series of \texttt{NB\_Igatherv} calls on each buffer.

\item \textbf{Auxiliary Operators}:
\cylon\ supports all auxiliary operators discussed in Section \ref{sec:model}. 
These operators are implemented with utilities developed in-house and from Arrow Compute, and for \gcylon{}, we use CuDF utilities where applicable. 

\item \textbf{Distributed Operators}
Except for \textit{Window} and \textit{Transpose}, \cylon\ implements the rest of the operators identified in Table \ref{tab:modin-ops}. As shown in Figure \ref{fig:dist-op}, all of them are implemented as a composition of local, auxiliary and communication operators based on the aforementioned patterns. Currently the \texttt{pandas} operator coverage is at a moderate 25\%, and we are working on improving the coverage.




\end{itemize}

\subsection{Features}
\label{sec:features}


\begin{itemize}[wide=0pt]
    \item  \textbf{Scalability and High Performance}: \cylon\ achieves above-average scalability and higher performance than the commonly used distributed dataframe systems. In Section \ref{sec:exper}, we compare strong scaling of \cylon{}, Modin, and Dask. 


\item \textbf{Flexible Dataframe API}: 
\cylon\ API clearly distinguishes between local and distributed operators with minimal changes to the \texttt{pandas} API semantics. This allows complex data manipulations for advanced users. As an example, a \texttt{join} (shuffle) can be easily transformed into a \texttt{broadcast\_join} just by changing a few lines of code.

\begin{lstlisting}[basicstyle=\scriptsize,language=Python]
df1 = read_csv_dist(..., env) # large df 
df2 = read_csv(...) if env.rank == 0 else None # read small df at rank 0
df2_b = env.broadcast(df2, root=0) # broadcast 
df3 = df1.merge(df2_b, ...) # local join
\end{lstlisting}


\item \textbf{Extensibility}: With the proposed model, \cylon\ was able to switch between multiple communication frameworks. Additionally, we extended this model to develop an experimental distributed memory dataframe for GPUs, \gcylon\, with minimum development effort.


\end{itemize}

\section{Experiments}
\label{sec:exper}

Our experiments were carried out in a 15-node Intel\textsuperscript{\textregistered} Xeon\textsuperscript{\textregistered} Platinum 8160 cluster. Each node has a total RAM of 255GB, uses SSD for storage and are connected via Infiniband with 40Gbps bandwidth. A maximum of 40 (of 48) cores were used from each node. 
The software used: Python v3.8 \& Pandas v1.4; \cylon\ (GCC v9.4, OpenMPI v4.1, \& Apache Arrow v5.0); Modin v0.12 (Ray v1.9); Dask v2022.1.
Uniformly random distributed data was used with two \texttt{int64} columns, $10^9$ rows ($\sim$16GB), and $\mathbf{C}=0.9$. This constitutes a worse-case scenario for key-based operators. 
The scripts to run these experiments are available in Github \cite{cylon-exp}.

The main goal of these operator benchmarks was to show how such generic patterns helped \cylon\ achieve scalable high performance. 
Dask and Modin operators are compared here only as a baseline. 
We tried our best to refer to publicly available documentation, user guides and forums while carrying out these tests to get the optimal configurations. 

\begin{itemize}[wide=0pt]
    \item \textbf{Scalability}: 
Figure \ref{fig:sperf} depicts strong scaling for the patterns. Dotted lines represent the speed-up over \texttt{pandas} ($pandas\_time/time$). Compared to Dask, Modin, and \texttt{pandas}, \cylon\ shows consistent performance and superior scalability. When the parallelism is increased from 1 to 256, the wall-clock time is reduced. 
Operations takes longer to complete at 512 parallelism. Per executor work is at its lowest in this instance, therefore the communication cost dominates over computation. 
For EP, a \textit{Barrier} is called at the end and it might carry some communication overhead. \cylon's local operators also perform on par or better than \textit{pandas}, which validates our decision to develop in a C++ backend. 

\begin{figure}[htbp]
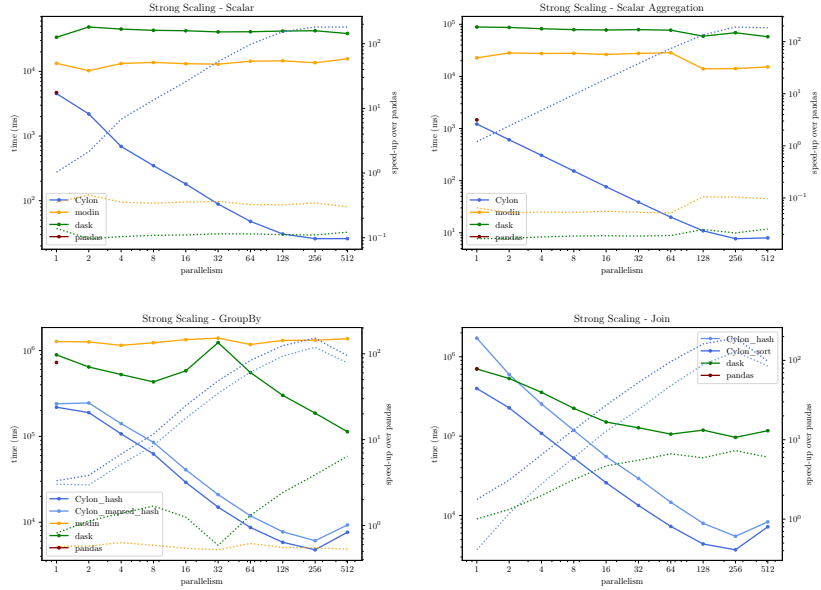

\begin{tabular}[width=\textwidth]{cc}
\scriptsize
\resizebox{0.45\textwidth}{0.33\height}{\input{"fig/plots/scalar.pgf"}} & 
\resizebox{0.45\textwidth}{0.33\height}{\input{"fig/plots/scalar-agg.pgf"}} \\
\resizebox{0.45\textwidth}{0.33\height}{\input{"fig/plots/groupby.pgf"}} & 
\resizebox{0.45\textwidth}{0.33\height}{\input{"fig/plots/join1B.pgf"}} 
\end{tabular} 
\captionsetup{justification=centering,skip=0pt}
\caption{Strong Scaling (1B rows, Log-Log) with speed-up over  \texttt{pandas}}
\label{fig:sperf}
\end{figure}
Unfortunately, Modin \texttt{join} for 1B rows failed, therefore we ran a smaller 100 million row test case (Figure. \ref{fig:sperf2}(a)). It only uses \texttt{broadcast-join} \cite{modin-issues}, which explains the lack of scalability. However, we encountered similar problems for the rest of the operators (Figure \ref{fig:sperf}).
Compared to Modin, Dask showed comparable scaling to \cylon\ for both \texttt{groupby} and \texttt{join}. Still it is disconcerting not to see any speed-up for both scalar tests.
\begin{figure}[htbp]
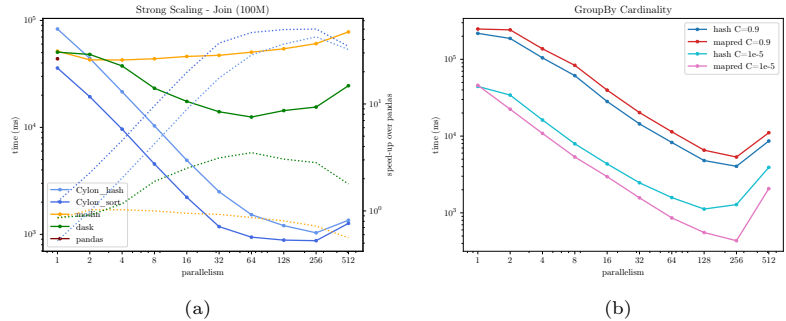

\begin{tabular}[width=\textwidth]{cc}
\resizebox{0.45\textwidth}{0.33\height}{\input{"fig/plots/join100M.pgf"}} & 
\resizebox{0.45\textwidth}{0.33\height}{\input{"fig/plots/cylon_groupby.pgf"}} \\
\scriptsize (a) & \scriptsize (b)
\end{tabular} 
\captionsetup{justification=centering,skip=0pt}
\caption{a: Strong Scaling Joins with Modin (100M rows, Log-Log),\\
b: Cardinality Impact on Combine-Shuffle-Reduce (\texttt{groupby}, 1B rows, Log-Log)}
\label{fig:sperf2}
\end{figure}


\item \textbf{Cardinality Impact}:
Figure \ref{fig:sperf2}(b) illustrates the impact of Cardinality ($\mathbf{C}$) on the \texttt{groupby} performance. 
When $\mathbf{C}=0.9$, hash-groupby (shuffle-compute) consistently outperforms the mapred-groupby (combine-shuffle-reduce), because the local combining step does not reduce the shuffle workload sufficiently. Whereas when $\mathbf{C}=10^{-5}$, shuffled intermediate result size is significantly lesser, and therefore the latter is much faster. This shows that the same operator might need to implement several patterns and choose an implementation based on runtime characteristics. 

\end{itemize}






\section{Related Work}

Dask distributed dataframe\cite{rocklin2015dask} was the first and foremost distributed dataframe system. 
It was targeted at providing better performance in personal workstations. RAPIDS CuDF, later extended Dask DDF for GPU dataframes.  
In large-scale supercomputing environments, HPC-based systems like MPI (Message Passing Interface) \cite{MPI-3-0_2012}, PGAS (partitioned global address space)\cite{zheng2014upc++}, OpenMP, etc. performed better compared to Apache Spark\cite{apache-spark} and Dask \cite{kamburugamuve2018anatomy,wickramasinghe2019twister2,abeykoon2019streaming}). Modin \cite{petersohn2020towards}, Dask \cite{rocklin2015dask}, and Koalas (Apache Spark) are some of the emerging distributed dataframe solutions, but the domain shows a lot more room for improvement. HPC-based distributed data engineering systems show promising support for workloads running in supercomputing environments \cite{widanage2020high,abeykoon2020data,abeykoon2021hptmt,perera2020fast}, and this is the main motivation for this paper.

\section{Limitations \& Future Work}
\label{sec:fut}

\cylon\ \textit{Sort} and \textit{Window} operators are still under development. 
Additionally, larger scale experiments have been planned to provide more finer-grained analysis on communication and computation performance. 
\cylon\ execution currently requires dedicated resource allocation, which may be a bottleneck in a multi-tenant cloud environment. Furthermore, fault tolerance is another feature that is yet to be added. We believe that both BSP and asynchronous executions are important for complex data engineering pipelines and are currently working on integrating \cylon{} with Parsl \cite{babuji2018parsl} and Ray \cite{moritz2018ray}. This would enable the creation of individual workflows that run on BSP, each of which can be scheduled asynchronously, that would optimize resource allocation without hindering the overall performance.

\section{Conclusion}

We recognize that today's data science community requires scalable solutions to meet their ever-growing data demand. Dataframes are at the heart of such applications, and in this paper we proposed a framework based on a set of generic operator patterns that lays the foundation for building scalable high performance dataframe systems. We discussed how this framework complements the existing literature available. We also presented \cylon, a reference runtime developed based on these concepts and showcased the scalability of its operators against leading dataframe solutions available today. We believe that there is far more room for development in domain, and we hope our work contributes to the next generation of distributed dataframe systems.



\bibliographystyle{splncs04}
\bibliography{ref.bib}

\end{document}